\documentclass[journal=aamick,manuscript=article,reqno]{amsart}
\usepackage[version=3]{mhchem} 
\usepackage[T1]{fontenc}       
\usepackage[super]{natbib}
\usepackage{physics}
\calclayout
\usepackage[foot]{amsaddr}
\usepackage{subfig}
\usepackage{mathtools}
\usepackage{amssymb}
\usepackage{gensymb}
\usepackage{graphicx}
\usepackage{filecontents}
\usepackage{bm}
\usepackage{xr}

\title{The thermodynamic principle determining the interface temperature during phase change}

\author{Tom Y. Zhao$^\dagger$}

\author{Neelesh A. Patankar$^{\dagger,*}$}
\address[$^\dagger$]{Northwestern University, Department of Mechanical Engineering: 2145 Sheridan Road, Evanston, Illinois 60208, USA}
\address[$^*$]{E-mail: n-patankar@northwestern.edu}

\begin{document}

\maketitle
\begin{abstract}
\normalsize What is the interface temperature during phase transition (for instance, from liquid to vapor)? This question remains fundamentally unresolved. In the modeling of heat transfer problems with no phase change, the temperature and heat flux continuity conditions lead to the interface temperature. However, in problems with phase change, the heat flux condition is used to determine the amount of mass changing phase. This makes the interface temperature indeterminate unless an additional condition is imposed. A common approach in the modeling of boiling is to assume that the interface attains the saturation temperature according some measure of pressure at the interface. This assumption is usually applied even under highly non-equilibrium scenarios where significant temperature gradients and mass transport occur across the interface. In this work, an ab-initio thermodynamic principle is introduced based on the entropy production at the interface that fully specifies the associated temperature under non-equilibrium scenarios. Physically, the thermodynamic principle provides a theoretical limit on the space of possible phase change rates that can occur by associating the mass flux with a corresponding interfacial entropy production rate; a stronger statement is made that a system with sufficient degrees of freedom selects the maximum entropy production, giving the observed phase change rate and associated interface properties. This entropic principle captures experimental and computational values of the interface temperature that can deviate by over $50\%$ from the assumed saturation values. It also accounts for temperature jumps (discontinuities) at the interface whose difference can exceed $15 \degree C$. This thermodynamic principle is found to appropriately complete the phase change problem.
\end{abstract}

\section{Introduction}

In phase transition (e.g. liquid to vapor), the fundamental principle that dictates the temperature at the interface between the two phases has been debated and it remains an open question. 

The interface temperature is presumably intrinsically connected to the rate of phase change as well as the interface velocity. This is critical information to understand and model phase change heat transfer in variety of heat exchange applications. This information is also crucial in designing water purification processes like membrane distillation,\cite{2012MD} energy storage systems using latent heat batteries,\cite{2018Sarbu}, additive manufacturing techniques involving molten metal jets \cite{2019Simonelli}, and phase change memory technologies for nonvolatile solid state storage.\cite{2010Burr} 

Theoretical and computational models typically assume that the interface between the two phases attains the saturation temperature.\cite{1998Juric,2005Dhir} Experimental work using thermocouples with thicknesses on the order of microns \cite{2007Badam, 2017Gatapova} have resolved interface temperatures that are found to deviate significantly from the saturation assumption.\cite{1998Juric} Theoretical attempts to find a different interface condition, to replace the saturation temperature condition, include the kinetic theory \cite{1991Tanasawa, 1998Juric} and the statistical rate theory.\cite{2001Ward}

Kinetic theory expresses the entropy generation at the interface using a constitutive relationship with the parameter $\phi$ representing the kinetic mobility, or the relative strength of molecular attachment to a surface. However, the evaluation of $\phi$ requires an empirical evaporation coefficient $\alpha$, which is difficult to measure and can deviate by over three orders of magnitude from the theoretical value of unity.\cite{1998Juric, 2007Badam} The kinetic theory also underestimates the temperature jump measured in experiment by 3 to 4 orders of magnitude.\cite{2017Gatapova}

The statistical rate theory uses quantum mechanics to describe a relationship between the rate of phase change and the change in entropy associated with a molecule transferring from the liquid to the vapor phase.\cite{2001Ward} After measuring the interface properties (including temperatures) of the liquid and vapor side from experiment, the mass flux from phase change can be calculated based upon the material properties of the fluid, the molecular vibrational frequencies and the partition function for the fluid molecule. From a computational standpoint or generally in scenarios where the interface temperatures and properties are not known a priori, the rate of phase change cannot be obtained via this method and vice versa.   

In this work, we determine the thermodynamic relationship between the temperatures of both phases at the interface and the rate of interfacial entropy production $\sigma$. This mapping provides a theoretical limit on the space of possible interface temperatures and phase change rates; there exists a maximum rate of entropy production due to the competition or inverse relationship between the entropy jump from phase change and the heat flux carried away from the interface. This space of possible $\sigma$ is bounded from below by the second law of thermodynamics $\sigma\geq0$. 

Finally, we propose a stronger thermodynamic principle that fully determines the interface temperatures during the time evolution of a phase change system. It is found that the interface temperatures which maximize the entropy production rate $\sigma$ capture the full range of both experimental and computational data on interfacial properties of different fluids and solids under evaporation, condensation, and freezing processes. This thermodynamic principle prefaced on the maximum rate of entropy production \cite{2003Dewar,2010Niven} also determines the rate of phase change as a function of material properties and temperature boundary conditions in the far field; properties and field variables at the interface are not known or fixed a priori. 

The proposed entropy condition closes the formerly incomplete problem of phase change under nonequilibrium scenarios.

\section{The missing interface condition}
\subsection{The complete problem without phase change}
\begin{figure}[]
\centering
\includegraphics[width=8.5cm]{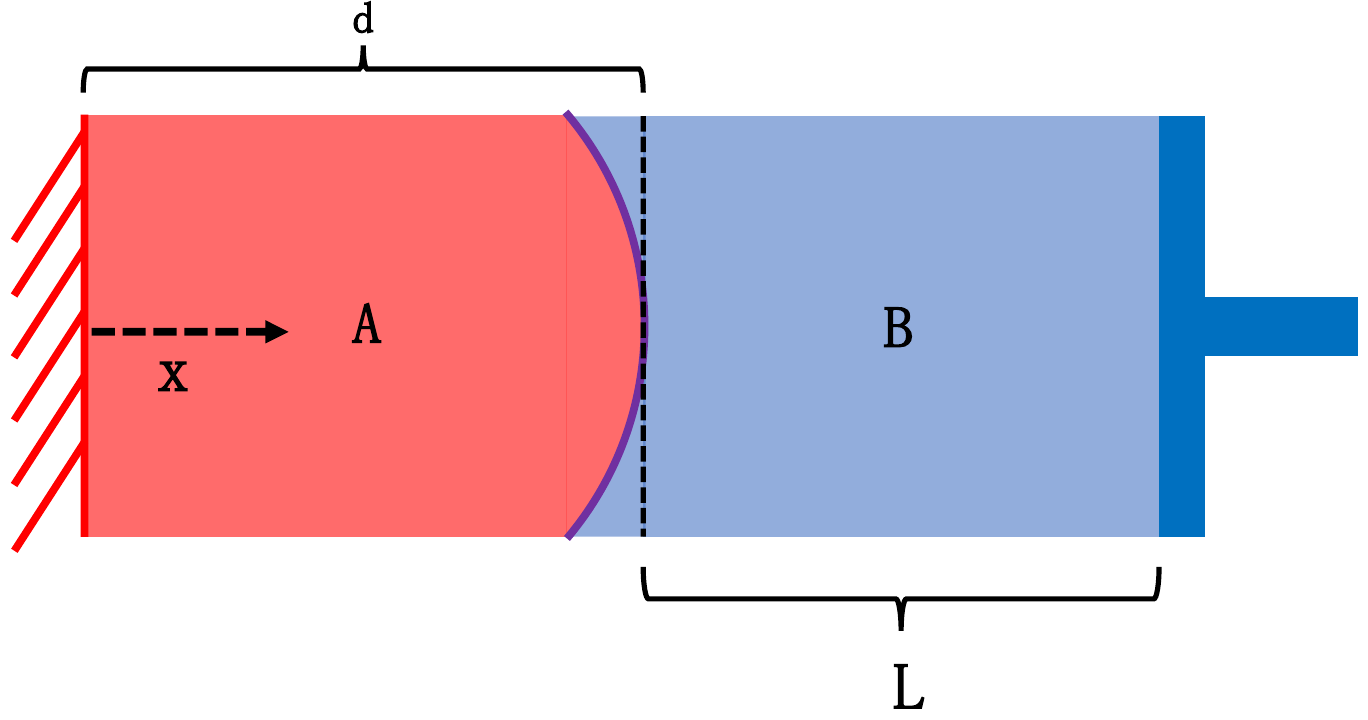}
\caption{Finite, one-dimensional system with two fluid components $A$ and $B $. The interface between the two phases is located at $x=d(t)$. A motionless wall bounds the domain at $x=0$, and a moveable piston at $x=L(t)+d(t)$ controls the pressure in the system. The wall is held at constant temperature $T_W$ and the piston at temperature $T_P$. The radius of curvature of the interface $R$ can be nonzero.}
\label{fig: problem_setup}
\end{figure}
We first consider a well-posed problem comprising a two component system in which no phase change can occur (for instance, with water and oil).  Figure \ref{fig: problem_setup}  shows a finite, one-dimensional system where the number of equations and boundary conditions can be easily counted. The governing equations for the incompressible species $A$ and $B$ across the two phase interface are

\begin{equation}
\pdv{u_c}{x}=0,
\label{eqn: cont}
\end{equation}
\begin{equation}
\rho_c\pdv{u_c}{t}=-\pdv{p_c}{x},
\label{eqn: mom}
\end{equation}
\begin{equation}
\rho_cc_{p,c}\pdv{T_c}{t}+\rho_cc_{p,c}u_c\pdv{T_c}{x}=k_c\pdv[2]{T_c}{x},
\label{eqn: energy}
\end{equation}
where the subscript $c\in[A,B]$. The velocities, pressures and temperatures in each phase are denoted $u_c$, $p_c$ and $T_c$ respectively. Similarly, $k_c$, $\rho_c$ and $c_{p,c}$ refer to the thermal conductivity, density and specific heat capacity at constant pressure. The boundary conditions at the motionless wall are
\begin{equation}
\text{at} \; x=0, \quad u_A=0,
\label{eqn: wallu}
\end{equation}
\begin{equation}
\quad \quad \quad \quad \quad \quad T_A=T_W.
\label{eqn: wallT}
\end{equation}
The boundary conditions at the piston are
\begin{equation}
\text{at} \; x=L(t), \quad p_B=p_{P},
\label{eqn: pistonp}
\end{equation}
\begin{equation}
\quad \quad \quad \quad \quad  \quad  T_B=T_{P}.
\label{eqn: pistonT}
\end{equation}
The interface conditions are
\begin{equation}
\hspace*{-0.6cm}\text{at} \; x=d(t), \quad \rho_A\bigl(u_A(d)-u_S\bigr)=\rho_B\bigl(u_B(d)-u_S\bigr),
\label{eqn: intu}
\end{equation}
\begin{equation}
\hspace*{-0.75cm}\rho_A\bigl(u_A(d)-u_S\bigr)=0,
\label{eqn: intu2}
\end{equation}
\begin{equation}
\quad \quad  [Q]= -k_B\pdv{T_B}{x}+k_A\pdv{T_A}{x}=0,
\label{eqn: intQ}
\end{equation}
\begin{equation}
\quad \quad T_A(d)=T_{SA}=T_{SB}=T_B(d),
\label{eqn: intT}
\end{equation}
\begin{equation}
\hspace*{-1.5cm} p_A=p_B+\gamma\kappa,
\label{eqn: intpress}
\end{equation}
where $u_S$ is the interface velocity, $T_W$ is the wall temperature, $T_{P}$ is the piston temperature, $p_{P}$ is the piston pressure, and $[Q]$ is the jump in heat flux across the interface, $\gamma$ is the surface tension between the two species and $\kappa$ is the interface curvature ($\kappa=1/R$ in one dimension). The temperatures $T_{SA}$, $T_{SB}$ represent the respective values for each species $A,B$ at the two-phase interface. Note that for simplicity, we have made the assumptions that the interface is massless, surface tension is constant and there is no temperature jump across the interface \cite{1998Juric}.

There are 6 unknown field variables ($u_A$, $T_A$, $p_A$, $u_B$, $T_B$, $p_B$) and six sets of conservation equations (eqn. \ref{eqn: cont}, \ref{eqn: mom}, \ref{eqn: energy}) for components A and B. The four mass and momentum conservation equations (eqns. \ref{eqn: cont} and \ref{eqn: mom} for components A and B) are first order differential equations that each require a single boundary condition (eqn. \ref{eqn: wallu}, \ref{eqn: pistonp}, \ref{eqn: intu}, \ref{eqn: intpress}). The two energy equations (eqn. \ref{eqn: wallT} for components A and B) are second order differential equations that each require two boundary conditions (eqn. \ref{eqn: wallT}, \ref{eqn: pistonT}, \ref{eqn: intQ}, \ref{eqn: intT}). Without phase change, the mass continuity condition at the interface (eqn. \ref{eqn: intu2}) specifies the interface velocity $u_S=u_A(d)=u_B(d)$ and completes the problem.
\subsection{The ill-posed problem with phase change}
Consider a scenario where phase change occurs between the two species (for instance, with liquid water and water vapor). The governing equations are the same, but the interface conditions change \cite{1992Carey, 1998Juric, 1974Delhaye}. Conservation of mass at the interface states
\begin{equation}
\text{at} \; x=d(t), \quad \rho_A\bigl(u_A(d)-u_S\bigr)=\rho_B\bigl(u_B(d)-u_S\bigr).
\label{eqn: intupc}
\end{equation}
From energy balance at the interface, the mass flux due to phase change is given as
\begin{equation}
u_B-u_A=\bigl(1/\rho_B-1/\rho_A\bigr)\frac{[Q]}{\Delta H}= \left(-k_B\pdv{T_B}{x}+k_A\pdv{T_A}{x}\right)\left(\frac{1/\rho_B-1/\rho_A}{\Delta H}\right).
\label{eqn: intQpc}
\end{equation}
For simplicity, the temperatures of the two phases at the interface are often assumed to be continuous, such that
\begin{equation}
T_A(d)=T_{SA}=T_{SB}=T_B(d).
\label{eqn: intTpc}
\end{equation}
Momentum conservation at the interface gives
\begin{equation}
\quad p_A=p_B+\gamma\kappa+(1/\rho_A-1/\rho_B)(\frac{[Q]}{\Delta H})^2,
\label{eqn: intpresspc}
\end{equation}
where viscous terms are neglected in the momentum conservation eqn. \ref{eqn: intpresspc} in the 1D limit. Here, $\Delta H=h_A-h_B$ is the latent heat of phase change expressed as the difference between the enthalpy of phases A and B, $h_A$ and $h_B$ respectively. 

As the mass conservation eqn. \ref{eqn: intupc} is no longer identically equal to zero due to the possible phase change between species A and B, the interface velocity $u_S$ becomes unspecified. The energy conservation condition eqn. \ref{eqn: intQpc} at the interface can be borrowed to fix the value of $u_S$, since the mass flux is determined by the thermal energy diffused to the interface relative to the latent heat of phase change. 

However, once the heat flux interface condition is used to determine the rate of phase change, the interface temperature $T_S$ becomes unspecified. Thus, the problem becomes ill-posed due to either the missing interface velocity $u_S$ or temperature $T_S$.
\section{The entropy condition at the two-phase interface}
The entropy condition at the two phase interface has been explored in the literature \cite{1974Delhaye}, but the resulting statement on the interfacial entropy production rate $\sigma$ (in units of energy per unit time and per unit area) is weak when referencing the second law of thermodynamics in or near equilibrium: $\sigma\geq 0$. We refer to this inequality as weak in that it is not sufficient to specify a particular interface temperature or velocity. Additionally, the statement of the second law leaves the entropy production rate unbounded in a semi-infinite range.

In this section, we will first present without reference to proposed laws or lemmas in nonequilibrium thermodynamics the entropy production rate across a two phase interface. Then for the simplified case of phase change across an interface without a temperature jump, it can be physically shown that the entropy production term is bounded from above by a maximum value. The range for $\sigma$ becomes finite, leading to a stronger statement on the possible macrostates accessible to the system.  

The rate of entropy production $\sigma$ at the two phase interface is given in general by \cite{1974Delhaye}
\begin{align}
\begin{split}
(T_S)\sigma=\dot{m}_A\biggl((T_{SA}-T_S)s_A+g_A-g_S+0.5(\bold{v}_A^2-2\bold{v}_A\cdot \bold{v}_S+\bold{v}_S^2)\biggr)\\+\dot{m}_B\biggl((T_{SB}-T_S)s_B+g_B-g_S+0.5(\bold{v}_B^2-2\bold{v}_B\cdot \bold{v}_S+\bold{v}_S^2)\biggr)\\+\bold{q}_A\cdot\hat{n}_A(1-T_S/T_{SA})+\bold{q}_B\cdot\hat{n}_B(1-T_S/T_{SB})+T_S\bold{q}_S\cdot\nabla_S(1/T_S)\\-\frac{\dot{m}_A}{\rho_A}(\tau_A\cdot\hat{n}_A)\cdot\hat{n}_A-\frac{\dot{m}_B}{\rho_B}(\tau_B\cdot\hat{n}_B)\cdot\hat{n}_B,
\label{eqn: fullent_Tjump}
\end{split}
\end{align}
where $T_{SA}$ and $T_{SB}$ are the absolute temperatures of the respective phases at the interface, and $T_S$ is the absolute interface temperature. Similarly, the variables $g_{A,B,S}$ are the free enthalpies, $s_{A,B}$ are the entropies per unit mass and $q_{A,B,S}$ are the heat fluxes. The unit normal vectors $\hat{n}_A$ and $\hat{n}_B$ are directed towards the interface; this convention agrees with the concept of a local entropy source at the interface \cite{1974Delhaye, 1973Meinhold}. The stress tensors in each species are given by $\tau_A$ and $\tau_B$, while the surface gradient is represented by $\nabla_S$. Finally, the mass fluxes across the interface are denoted as $\dot{m}_A = \rho_A(\bold{v}_A-\bold{v}_S)\cdot\hat{n}_A$ and $\dot{m}_B=\rho_B(\bold{v}_B-\bold{v}_S) \cdot\hat{n}_B$. Eqn. \ref{eqn: fullent_Tjump} for the entropy source term comes from combining the evolution equations for mass, momentum, energy and entropy at the interface \cite{1974Delhaye}.

Next, we can find a simplified expression for the entropy production rate in a 1D system across a massless, infinitesimally thin interface (Fig. \ref{fig: problem_setup}). Let $c\in[A,B]$. By definition, the sum of the free enthalpy $g_c$ and the product of temperature with the entropy density $T_{Sc}s_c$ of each phase is simply the saturation enthalpy, since the temperature and pressure dependencies of the two terms cancel to give
\begin{equation}
g_c+T_{Sc}s_c=g_{c,\text{sat}}+T_{Sc}s_{c,\text{sat}}(T_{Sc})=h_{c,\text{sat}}.
\label{eqn: gsh}
\end{equation}
The mass flux across the interface due to phase change $\dot{m}$ is specified by the energy balance equation at the interface (eqn. \ref{eqn: intQpc}) with
\begin{equation}
\dot{m}=\frac{1}{h_{A,\text{sat}}-h_{B,\text{sat}}}(\bold{q}_A\cdot\hat{n}_A+\bold{q}_B\cdot\hat{n}_B)=-\dot{m}_A=\dot{m}_B.
\label{eqn: mdot}
\end{equation}
We can thus simplify the full expression for the entropy production rate at the two phase interface to
\begin{align}
\begin{split}
T_S\sigma_{1M}=-\dot{m}\biggl(-T_Ss_A\biggr)+\dot{m}\biggl(-T_Ss_B\biggr)+q_A(-T_S/T_{SA})-q_B(-T_S/T_{SB})-\frac{\dot{m}^3}{2}\left(\frac{1}{\rho_A^2}-\frac{1}{\rho^2_B}\right).
\label{eqn: 1dmassless}
\end{split}
\end{align}
In this 1D expression for the entropy production rate $\sigma_{1M}$, the dependence of the heat fluxes $q_c$ and mass flux $\dot{m}$ on the interfacial temperatures $T_S,T_{SA},T_{SB}$ can be estimated from the boundary conditions in the far field and the assumption of linear temperature profiles in each phase. On the other hand, the variation of each species' entropy density $s_c$ with interface temperature can only be specified after more information is known about the composition of each phase.

For the particular case of phase change between vapor (species V) and liquid (species L), the entropy densities can be expressed in terms of the pressure and temperature in each phase as
\begin{equation}
(T_S)s_V=(T_S)s_{V,\text{sat}}-RT_S\text{ln}\left(\frac{p_V}{p_{\text{sat}}(T_{SV})}\right),
\label{eqn: sv}
\end{equation}
\begin{equation}
(T_S)s_L=(T_S)s_{L,\text{sat}}-\frac{1}{\rho_L}\left(p_L-p_{\text{sat}}(T_{SL})\right),
\label{eqn: sl}
\end{equation}
where $\rho_L$ is the density of the liquid phase, $R$ is the specific gas constant and $p_{\text{sat}}$ is the saturation pressure associated with the interface temperature $T_{SV}$ or $T_{SL}$ of each phase.

Following the 1D formulation of $\sigma_{1M}$ (eqn. \ref{eqn: 1dmassless}), the entropy production rate at the interface of a vapor-liquid system becomes
\begin{align}
\begin{split}
\sigma_{LV}(T_{SV},T_{SL})=\dot{m}\biggl(s_{V,\text{sat}}-s_{L,\text{sat}}-R\text{ln}\left(\frac{p_V}{p_{\text{sat}}(T_{SV})}\right)+\frac{1}{T_{SL}\rho_L}\left(p_L-p_{\text{sat}}(T_{SL})\right)\biggr)\\+q_V(-1/T_{SV})-q_L(-1/T_{SL})-\frac{\dot{m}^3}{2T_S}\left(\frac{1}{\rho_V^2}-\frac{1}{\rho^2_L}\right).
\label{eqn: 1dmasslessLV}
\end{split}
\end{align}
This poses the question: what is a reasonable choice for $T_S$? For simplicity, we take $T_S\approx T_{SL}$ in this work on the basis that the Knudsen layer in the liquid phase is significantly smaller than that in the vapor \cite{2017Gatapova}.
\subsection{Physical insight into the existence of a maximum entropy production rate}
The entropy production rate $\sigma$ at the two phase interface is bounded below by the second law of thermodynamics. In a simplified 1D system, we now show that $\sigma$ is bounded above by a maximum, finite value. 

Consider the setup introduced in Fig. \ref{fig: problem_setup}, where component A is water vapor (V) and component B is liquid water (L). The wall at $x=0$ is superheated to temperature $T_W$, while the piston at $x=L$ is maintained at the saturation temperature $T_P=T_{\text{sat}}$ corresponding to the applied piston pressure $p=p_{P}$. 

We make further assumptions to simplify the analysis and provide physical intuition into the competing effects that drive the entropy production rate to achieve a finite maximum value. First, let us suppose that the temperature in each region is diffusion dominated, such that the profile for $T$ is linear in the vapor and liquid phases. Next, we take $T_S=T_{SV}=T_{SL}$, acknowledging that this should only hold in special instances such as when the system is in equilibrium. With this, the mass flux at the interface (eqn. \ref{eqn: mdot}) becomes
\begin{equation}
\dot{m}=\rho_L(u_S-u_L)=\frac{1}{\Delta H}\left(k_L\frac{T_{P}-T_S}{L}-k_V\frac{T_S-T_{W}}{d}\right).
\label{eqn: mdot}
\end{equation}
Meanwhile, the entropy production rate at the interface (eqn. \ref{eqn: 1dmasslessLV}) is simplified to
\begin{align}
\begin{split}
\sigma^\text{s}_{LV}(T_{S})=\dot{m}\biggl(s_V-s_L \biggr)-\frac{\dot{m}\Delta H}{T_S},
\label{eqn: simp1dmasslessLV}
\end{split}
\end{align}
where we have neglected the cubic term in mass flux, since $\dot{m}<<1$ is typically a good approximation. If the lengths $d$, $L$, the temperatures $T_P$ and $T_W$ as well as the piston pressure $p_P$ in the far field are both fixed, then $\sigma$ is only a function of $T_S$. 

As $T_S$ increases, the magnitude of $\dot{m}$ decreases along with the net heat transfer to the interface. Meanwhile, the difference $s_V-s_L$ decreases as well (eqn. \ref{eqn: sv} and \ref{eqn: sl}) with larger $T_S$ \cite{2012NIST}. Thus the first term in the simplified interfacial entropy evolution $\sigma^\text{s}_{LV}$ is inversely proportional to the interface temperature. 

The second term -$\frac{\dot{m}\Delta H}{T_S}$ is directly proportional to $T_S$. As the interface temperature increases, $\dot{m}$, $\Delta H$ and $\frac{1}{T_S}$ all decrease, such that the negative of their product increases. Due to the competition between the entropy flux due to phase change and the heat flux from the interface in eqn. \ref{eqn: simp1dmasslessLV}, $\sigma^\text{s}_{LV}$ reaches a maximum value with respect to $T_S$, analogous to how the change in Gibbs free energy goes through a minimum in the classical heterogeneous nucleation theory due to the competition between surface tension and volumetric free energy as the radius of the nucleus varies \cite{1992Carey,2020Zhao}.

\begin{figure}[]
\centering
\includegraphics[width=8.5cm]{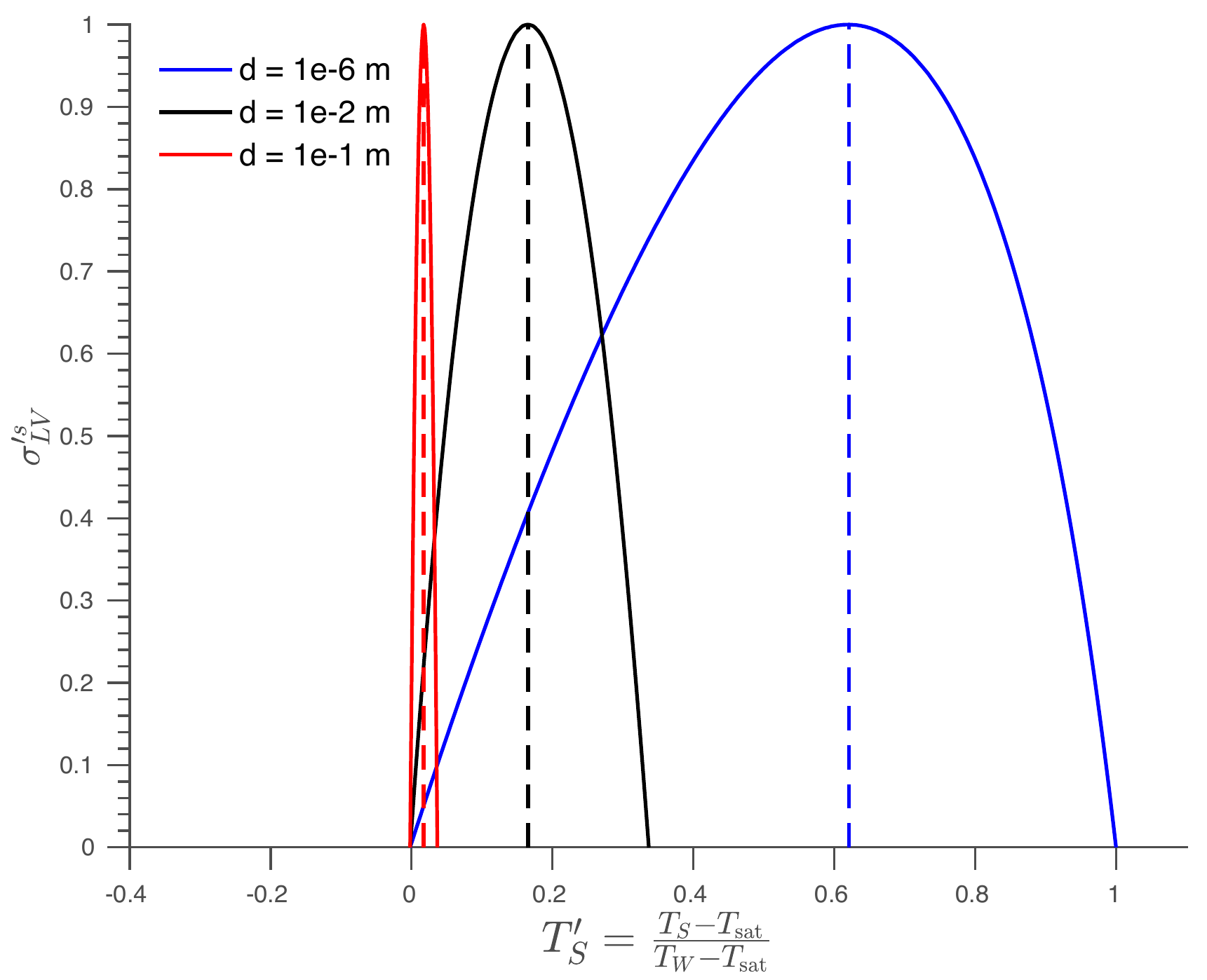}
\caption{The entropy evolution at the interface (eqn. \ref{eqn: simp1dmasslessLV}) normalized by the maximum value ($\sigma'^s_{LV}=\frac{\sigma^s_{LV}}{\max\limits_{T_S}(\sigma^s_{LV})}$) as a function of the normalized interface temperature $T'_S$. The entropy production rate $\sigma'_C$ reaches a maximum between $T'_S=1$ (the wall temperature) and $T'_S=0$ (the saturation temperature) due to opposing dependencies between the mass flux and the difference in phase free entropies on the interface temperature. The temperature corresponding to the maximum $\sigma'^s_{LS}$ drops towards the saturation value as the distance $d$ between the wall and interface increases to macroscale lengths. The vapor and liquid properties of water (thermal conductivities, latent heat of phase change, etc.) as a function of temperature were referenced from NIST\cite{2012NIST}. The boundary conditions and location were set to $L=0.1$ m, $T_{\text{wall}}=550$ K, $T_{\text{sat}}=373.15$ K, $p_{P}=1$ atm. Linear temperature profiles are assumed in the liquid and vapor domains.}
\label{fig: G_vs_T}
\end{figure}

Fig. \ref{fig: G_vs_T} shows that $\sigma^s_{LV}$ achieves a finite maximum value at an interface temperature $T^*_{S}$ satisfying $T_{\text{sat}}<T^*_{S}<T_{\text{wall}}$, in agreement with our qualitative analysis of the simplified entropy production rate. As the distance between the superheated wall and the liquid-vapor interface increases from micro to macro length scales, the interface temperature $T_{S}$ drops towards the saturation temperature corresponding to the applied piston pressure. This is accompanied by a shrinkage of
the temperature range where $\sigma_{C}\geq0$, suggesting that the traditional assumption of $T_{\text{sat}}$ at the interface is only valid at macroscale. 

\subsection{The maximum entropy principle at the two-phase interface}
Having demonstrated that the interfacial entropy production rate achieves a maximum, finite value for phase change systems, we now propose an entropic condition to determine the exact temperatures at the interface.

In nonequilibrium thermodynamics, it has been proposed that a process follows the path along which the entropy produced in the system at each step is maximized, subject to conservation laws as well as external constraints such as prescribed thermodynamic forces or fluxes \cite{2006MEP}. 

The maximum entropy production principle (MEPP) therefore seeks to generalize the inequality formulation of the second law of thermodynamics, which only states that the entropy production is either positive for irreversible processes or zero for reversible ones \cite{1974Delhaye}; alone, the second law gives a possible range of discontinuous interface temperatures $T_{SA}$ and $T_{SB}$ that satisfy $\sigma\geq0$, but does not pinpoint an actual value for $T_{SA}$, $T_{SB}$. 

MEPP has been explored with series of proofs in the literature ranging from variational analyses to statistical mechanics considerations \cite{2006MEP, 2010Niven, 2003Dewar, 2010Martyushev}. Functionally, MEPP can be used as a variational principle to solve the Boltzman equation \cite{2006MEP}; in climate models to predict surface temperatures and cloud coverage \cite{1978Paltridge}; in solid state physics to predict dendritic structure and growth rates \cite{2010Vladislav}. 

Here, we propose that the MEPP closes the phase change problem at the two phase interface, in that the maximum entropy production rate can be used to exactly specify the interface temperatures. In general, suppose that the discontinuous temperatures $T^*_{SA}$ and $T^*_{SB}$ give an optimal solution to max($\sigma(T_{SA},T_{SB})$) while satisfying the imposed constraints on thermodynamic fluxes or forces and conservation laws, represented by the series of conditions $F_i$($T_{SA},T_{SB}$)=0 for $i\in[1,2,...,N]$; in general, the dependence of $\sigma(T_{SA},T_{SB})$ on the interface temperatures is given by eqn. \ref{eqn: fullent_Tjump}. Then $T^*_{SA}$ and $T^*_{SB}$ are the temperatures observed at the two phase interface when the diffusive time scale $\tau_D$ of the system is smaller than the evolutionary time scale $\tau_E$. 

Here, we take $\tau_D=d_c^2/\alpha_c$ to be the thermal diffusion time scale such that $\alpha_c$ is the thermal diffusivity of phase $c\in[A,B]$ at the associated interface temperature and $d_c$ is the relevant length scale occupied by phase $c$. The evolutionary time scale $\tau_E=d_c/u_S$ is the length scale divided by the interface speed $u_S$, which likewise is a function of the interface temperature. We find that for systems that satisfy $\tau_D<\tau_E$, the interface temperatures and phase change rates predicted by MEPP capture the corresponding data obtained from experiment. 

The physical intuition behind this closure condition is that the trajectory of states corresponding to the maximum entropy production rate at each step reflects the most probable path observed in the system under the constraints imposed by fixed thermodynamic fluxes or forces as well as conservation laws \cite{2006MEP,2010Niven}. Endres showed that for the Schl\"ogl model of a first order phase transition with noise, the probability of observing a particular trajectory at nonequilibrium steady state increases exponentially with the entropy production rate\cite{2017Endres}. Specifically, the most probable trajectory of states is the one that maximizes the entropy production rate, while minimizing the classical and stochastic action along that path. The latter condition represents that the governing equations describing the system are satisfied.

\section{Results}
The interfacial temperatures that maximize the entropy production rate can be used to describe phase change features in both simulation and experiment. Only properties in the far field need to be know a priori in order to predict temperatures and mass fluxes at the interface. The efficacy of the proposed thermodynamic closure principle prompts its use in experiment and continuum simulation to capture the evolution of the interface under nonequilibrium behavior.
\subsection{Liquid-vapor interface temperature at nanoscale}
In this work, we use molecular dynamics (MD) simulations to explore nonequilibrium phase change across the liquid-vapor interface of water within nanometers of the superheated wall. The simulation setup in Fig. \ref{fig: problem_setup} is established across a distance $L+d$ of $30$ nm, with phase A adjacent to the superheated wall referring to water vapor and phase B adjacent to the piston referring to liquid water; further details are provided in the Methods section below.

\begin{figure}[]
\centering
\includegraphics[width=8.5cm]{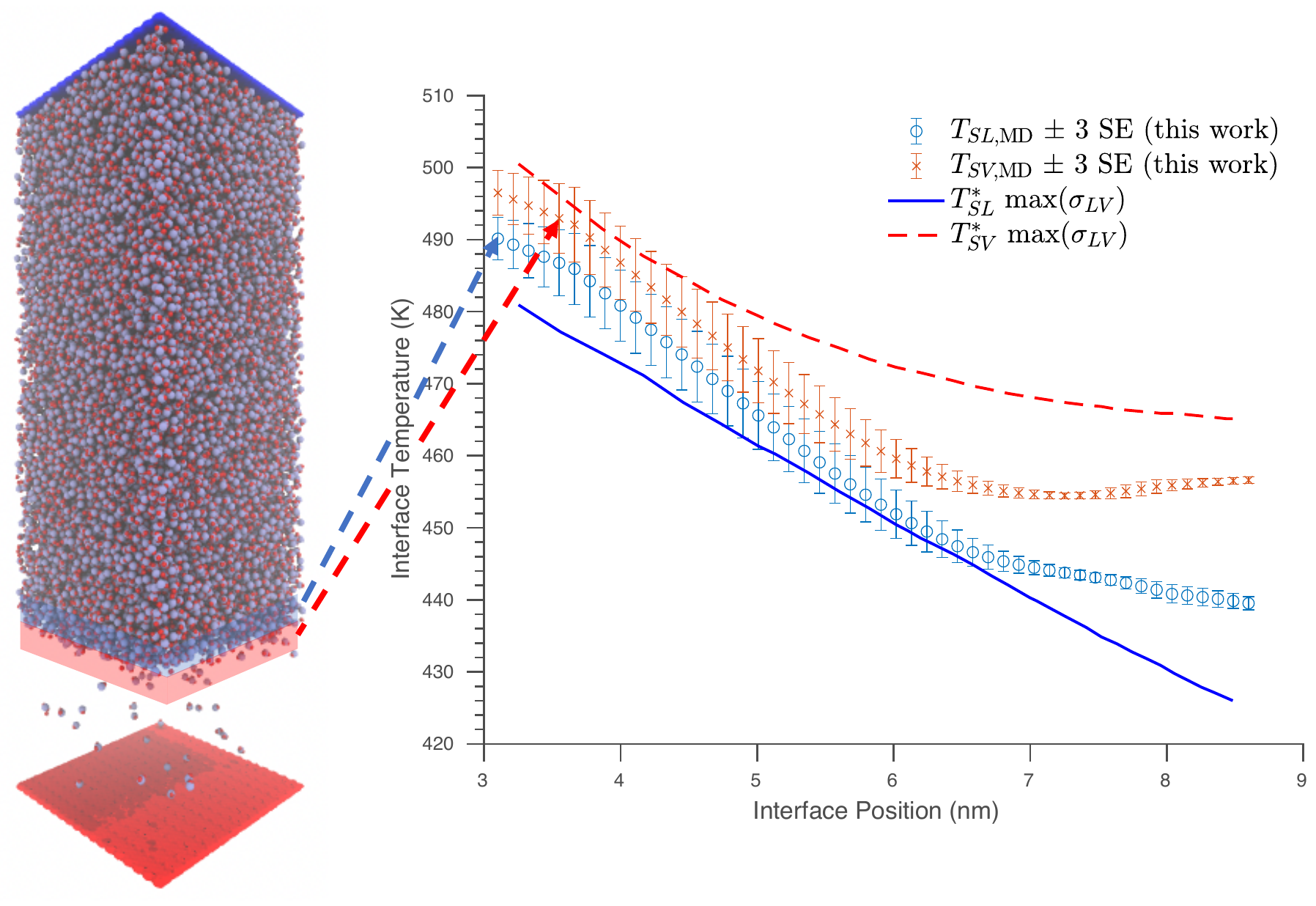}
\caption{The evolution of liquid and vapor side interface temperatures as a function of the interface distance $d$ from the superheated wall, predicted by the maximum entropy rate principle proposed in this work and measured via MD. The temperatures as averaged over 6 molecular dynamics simulations from independent initial conditions are well captured by maximizing the interfacial entropy production rate at each time step. The vapor and liquid properties of water (thermal conductivities, latent heat of phase change, etc.) as a function of temperature were referenced from NIST\cite{2012NIST}. The boundary conditions were $T_{W}=575$ K, $T_{P}=373.15$ K, $p_{P}=1$ bar. Linear temperature profiles were assumed in the liquid and vapor domains.}
\label{fig: MD_liqvap}
\end{figure}

Figure \ref{fig: MD_liqvap} demonstrates that the liquid and vapor side temperatures ($T^*_{SL}$ and $T^*_{SV}$) associated with the maximum entropy production rate at the two phase interface (eqn. \ref{eqn: 1dmasslessLV}) well captures the interfacial temperatures $T_{SL,\text{MD}}$ and $T_{SV,\text{MD}}$ measured using molecular dynamics as the system evolves during the phase change process. The constant saturation temperature $T_{\text{sat}}=373.15$ K corresponding to the pressure $p_P=1$ bar applied at the piston completely fails to model the non-constant, discontinuous dynamics of the interface temperatures.  

In the window of time presented in which the vapor film thickness $d$ increases from $3.5$ to $9$ nm, the two phase interface shifts away from the superheated wall mainly due to expansion of the superheated vapor. In fact, the average mass flux measured via MD when the interface position $3.5<d<9$ nm is $\overline{\dot{m}}=-414.1\frac{\text{kg}}{\text{m}^2\text{s}}$, which means that condensation occurs at the interface. The mass flux that maximizes the entropy production rate averaged over $3.5<d<9$ nm is $\overline{\dot{m}^*}=-478.1\frac{\text{kg}}{\text{m}^2\text{s}}$. Thus, this entropic interface principle accurately describes the physically unintuitive mode of condensation at the interface near a superheated vapor and estimates the correct order of magnitude of the phase change rate of a nonequilibrium, nanoscale process. 

Possible sources of error in this analysis include the deviation of associated fluid properties such as thermal conductivity or enthalpy from bulk values for the SPC/E water model compared to real water as well as for the nanoscale film compared to the bulk phase \cite{2020Zhao}. Additionally, it was assumed that the temperature profiles in both the liquid and vapor follow linear regimes.

In this system, as in all MD simulations and experimental outcomes included in this work, the thermal diffusion time scale $\tau_D$ is smaller than the evolutionary time scale of the interface $\tau_E$. Here, $\tau_{DL}\approx 1\mathrm{e}{-9}$ s in the liquid phase and $\tau_{DV}\approx 1\mathrm{e}{-11}$ s in the vapor phase are one and three orders of magnitude smaller than $\tau_E$, respectively. Intuitively, this means that the system has time to resolve interfacial temperatures that maximize the entropy production rate before the interface shifts to a new location, such that the interface can be considered stationary with respect to the entropy production rate. 

\subsection{Ice-liquid interface temperature at nanoscale}
\begin{figure}[]
\centering
\includegraphics[width=8.5cm]{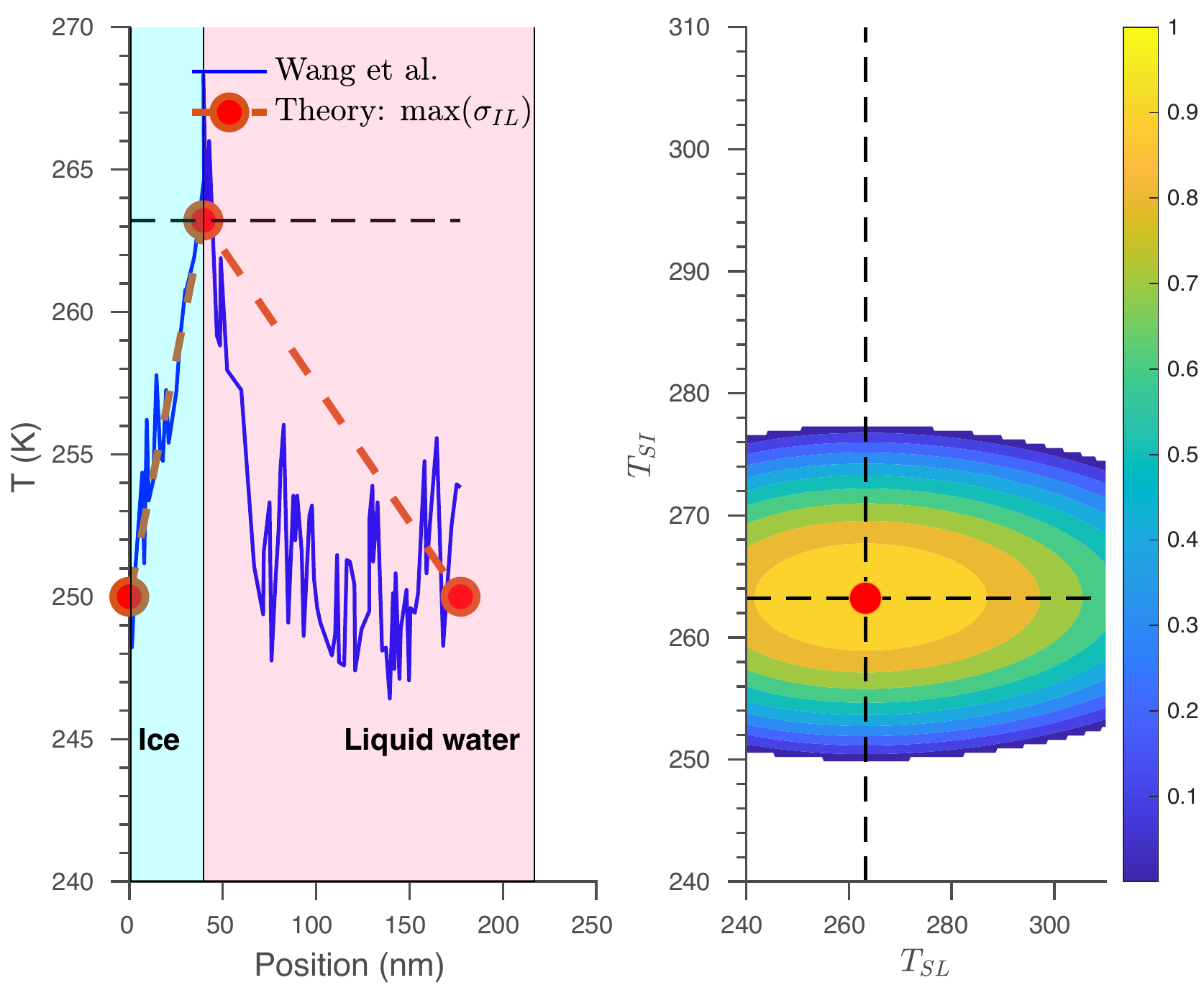}
\caption{\textbf{A}) The temperature profile of an ice-liquid system as measured via molecular dynamics by Wang et al\cite{2017Wang} using the mW water model and as calculated by the maximum entropy production interface condition proposed in this work. Note that molecular dynamics places the interface temperature $T_{SL,\text{MD}}\approx T_{SI,\text{MD}}\approx 265$ K to be $15$ K greater than the boundary temperatures $T_{BC}\approx250$ K but around $9$ K less than the freezing point of mW water $T_F=274.6$ K. The temperature distribution is non-monotonic and yet cannot be approximated accurately by the saturation temperature assumption. \textbf{B}) The interfacial entropy production rate exhibits a maximum value at $T^*_{SL}=T^*_{SI}=263.2$ K. The ice and liquid properties of water (thermal conductivities, latent heat of phase change, etc.) were drawn from the mW water properties reported by Wang et al. \cite{2017Wang}. Linear temperature profiles were assumed in the liquid and ice domains. }

\label{fig: MD_liqice}
\end{figure}

The agreement between the proposed entropy production interface principle and molecular dynamics simulation holds for phase change between solid and liquid water as well. Figure \ref{fig: MD_liqice} shows that the ice and liquid side interface temperatures $T^*_{SI}$ and $T^*_{SL}$ corresponding the maximum interfacial entropy production rate well describe the the highly non-equilibrium interface temperatures measured by MD during the freezing process. 

Note that the entropy production rate $\sigma_{IL}$ at the interface between the ice and liquid phases can be derived from the general 1D expression (eqn. \ref{eqn: 1dmassless}) by taking the pressure dependence of the ice phase entropy to be
\begin{equation}
(T_S)s_I=(T_S)s_{I,\text{sat}}-\frac{1}{\rho_I}\left(p_I-p_{\text{sat}}(T_{SI})\right).
\label{eqn: si}
\end{equation}
The interfacial entropy production rate $\sigma_{IL}$ can then be expressed as
\begin{align}
\begin{split}
\sigma_{LV}(T_{SI},T_{SL})=\dot{m}\biggl(s_{L,\text{sat}}-s_{I,\text{sat}}-\frac{1}{T_{SL}\rho_L}\left(p_L-p_{\text{sat}}(T_{SL})\right)+\frac{1}{T_{SI}\rho_I}\left(p_I-p_{\text{sat}}(T_{SI})\right)\biggr)\\+q_L(-1/T_{SL})-q_I(-1/T_{SI})-\frac{\dot{m}^3}{2T_S}\left(\frac{1}{\rho_L^2}-\frac{1}{\rho^2_I}\right),
\label{eqn: 1dmasslessIL}
\end{split}
\end{align}
where the difference in liquid and ice phase entropies $s_{L,\text{sat}}-s_{I,\text{sat}}$ for the mW water model was given by Holten et al\cite{Holten_2013}.

The interface temperatures appear approximately continuous ($T_{SI,\text{MD}}\approx T_{SL,\text{MD}}\approx T_{S,\text{MD}}\approx265$ K), but nonetheless form non-monotonic profiles with the boundary conditions in the liquid and ice domains $T_{BC}\approx250$ K. Thus, the simplest assumption of a constant temperature distribution in both regions $T_{S,\text{MD}}\approx T_{BC}$ fails. The freezing point of the mW water model $T_F=274.6$ K exceeds $T_{S,\text{MD}}$ by around $9$ K and is thus also a poor predictor.

 The best estimate for the interface temperature is obtained by maximizing the entropy produced at the ice-liquid interface while assuming linear temperature profiles in both the liquid and vapor domain. The interfacial velocity associated with this maximum entropy rate principle $u^*_{S}=4.37$ m/s also agrees with the interface velocity measured via molecular dynamics $u_{S,\text{MD}}=4.17$ m/s.

\subsection{Experiments at mesoscale} 
The stochastic rate theory (SRT) suggested by Ward et al. advanced meticulous experiments to measure the interfacial temperature jump between two phases \cite{1999Ward2, 2001Ward, 2008Ward, 1999Ward, 2004Ward, 2007Badam}. As noted, the SRT gives reasonable estimates for the temperature jump if the interface temperature on either the liquid ($T_{SL}$) or vapor ($T_{SV}$) side as well as the mass flux across the interface is measured first. 

\begin{figure}[]
\centering
\includegraphics[width=8.5cm]{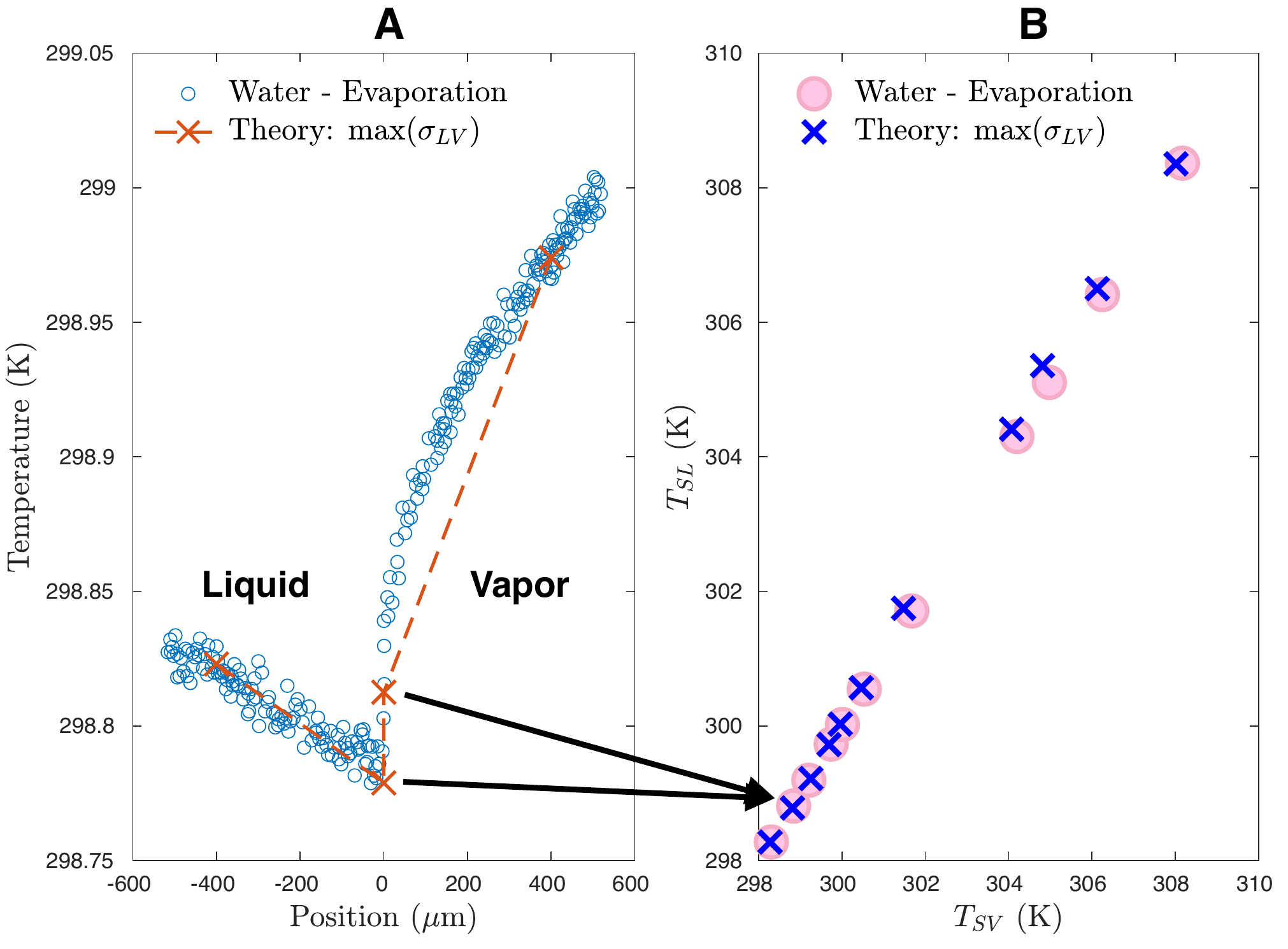}

\caption{\textbf{A}) The experimental temperature profile \cite{2017Gatapova} of a two phase water system is well described by the predicted interface temperatures for the liquid ($T_{SLM}$) and vapor ($T_{SVM}$) side using the maximum entropy principle. The interface is located at $x=0$ micron, where a pronounced temperature jump creates a non-monotonic temperature distribution such that $T_{SL}$ is not bounded by the liquid or vapor temperatures in the far field. \textbf{B}) The interfacial temperature jump is well described by the maximum entropy principle. Only data sets that provided all necessary information such as boundary conditions, distances to the interface, etc. were included in the plot to avoid using any unknown properties to 'fit' the data. Linear temperature profiles were assumed in the liquid and vapor domains. The vapor and liquid properties of water and octane (thermal conductivities, latent heat of phase change, etc.) as a function of temperature were referenced from the IAPWS formulation \cite{2012Huber, 2002Wagner, 2012NIST}.} 
\label{fig: liqvap}
\end{figure}

Fig. \ref{fig: liqvap} shows that the experimental interface temperatures measured via micro-thermocouples \cite{2017Gatapova} are well captured by the vapor and liquid interface temperatures $T^*_{SV}$, $T^*_{SL}$ that maximize $\sigma_{LV}$ (eqn. \ref{eqn: 1dmasslessLV}). Therefore in applications where interface properties are not available a priori, the maximum entropy production condition may be used to pinpoint the absolute temperatures of both phases at the interface and the mass flux due to phase change. Only far field properties such as temperature and pressure must be input into this analysis. 

The bulk temperature profiles in Fig. \ref{fig: liqvap}\textbf{A} are non-monotonic due to the interfacial temperature jump, and the liquid side interface temperature $T_{SL}$ is not bounded by the temperature conditions in the far field. This escapes a straightforward description from existing theory and heretofore falls under the umbrella of 'nonlinear, transient evolution'. However, the nonequilibrium thermodynamic mechanism proposed in this work suggests that the interfacial liquid and vapor temperatures are selected for by maximizing the rate of entropy produced due to phase change; this non-monotonic and "unbounded" behavior of the interface temperatures is therefore deterministic, rather than stochastic. 

\begin{figure}[]
\centering
\includegraphics[width=8.5cm]{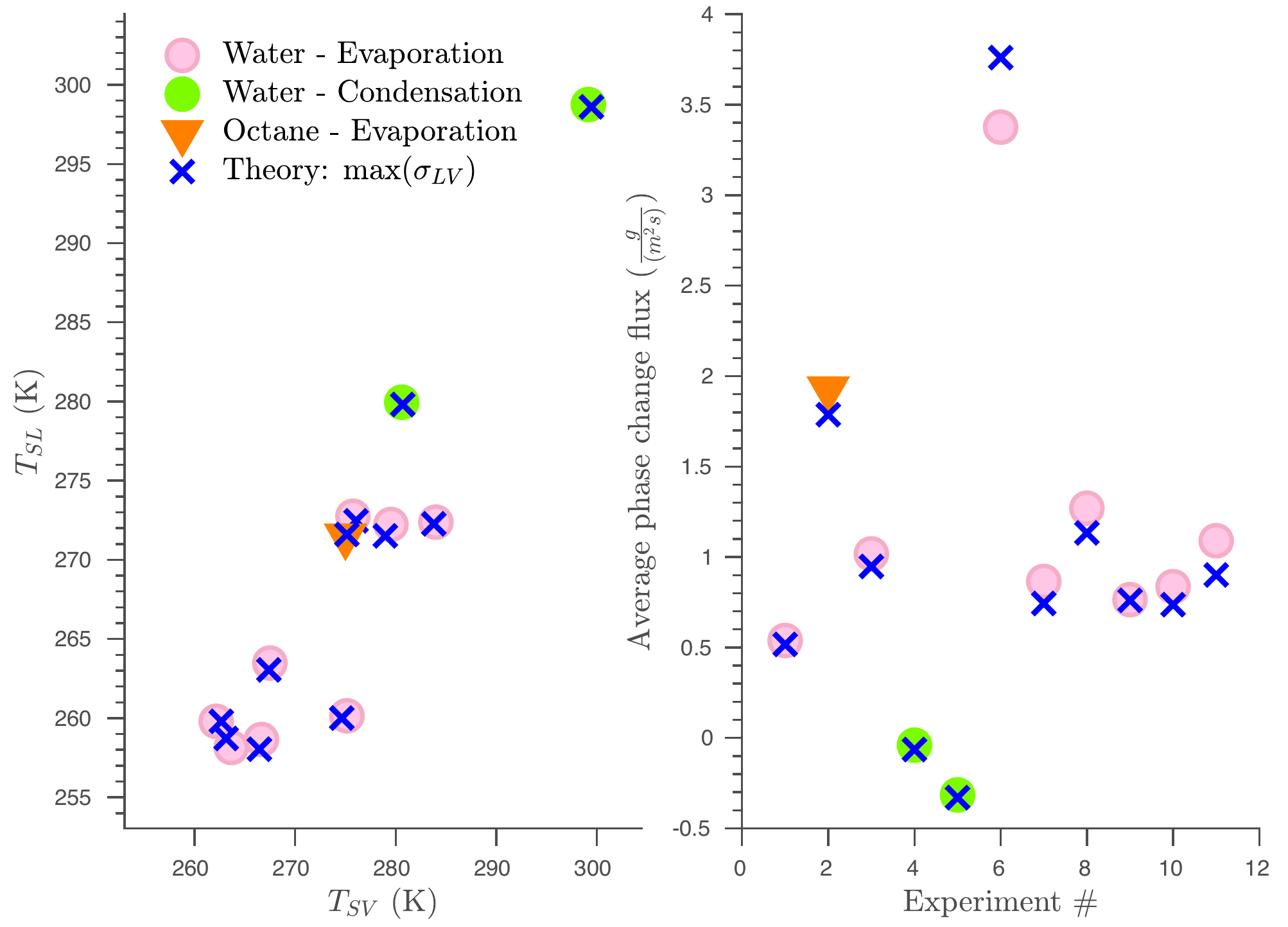}

\caption{\textbf{A}) The experimentally measured temperature jump as well as the distinct liquid and vapor side temperatures at the two-phase interface are captured by the maximum entropy principle. This general agreement between experiment and theory holds for evaporation (Exp $\#$ 1, 3, 7, 8, 9)\cite{1999Ward2, 2001Ward, 2008Ward} and condensation (Exp $\#$ 4, 5)\cite{2001Ward} of water; evaporation of octane (Exp $\#$ 2) \cite{1999Ward}; evaporation of water under turbulent conditions (Exp $\#$ 6) \cite{2004Ward}; and evaporation of water heated on the vapor side (Exp $\#$ 10, 11, 12) \cite{2007Badam}. \textbf{B}) The average phase change rate at the interface in units of mass per area, per unit time drawn from the same experiments are also captured by the maximum entropy principle. Only data sets that provided all necessary information such as boundary conditions, distances to the interface, etc. were included in the plot to avoid using any unknown properties to 'fit' the data; data points that overlapped significantly in the plot were also excluded for clear visualization. The vapor and liquid properties of water and octane (thermal conductivities, latent heat of phase change, etc.) as a function of temperature were referenced from the IAPWS formulation \cite{2012Huber, 2002Wagner, 2012NIST}. }

\label{fig: alljumps}
\end{figure}

Another set of experimental results\cite{1999Ward2, 2001Ward, 2008Ward,1999Ward,2007Badam} are visualized in Fig. \ref{fig: alljumps}. The temperatures of each phase at the interface in Fig. \ref{fig: alljumps}\textbf{A} and mass fluxes due to phase change in Fig. \ref{fig: alljumps}\textbf{B} are well described by the thermodynamic principle proposed in this work. This agreement holds for evaporation and condensation of water under laminar and turbulent conditions, as well as for evaporation of octane. Although temperature jumps in these sets of experiments are typically smaller, the nonequilibrium phase change processes examined nonetheless exhibit the distinct unbounded characteristic wherein the interface temperatures can lie outside the range of the far field temperature conditions. In all cases, only far field pressures and temperatures were used as input into the maximum entropy rate principle; all properties on the interface were determined by maximizing the entropy produced $\sigma_{LV}$ (eqn. \ref{eqn: 1dmasslessLV}).  
 
\section{Discussion}
\begin{figure}[]
\centering
\includegraphics[width=10cm]{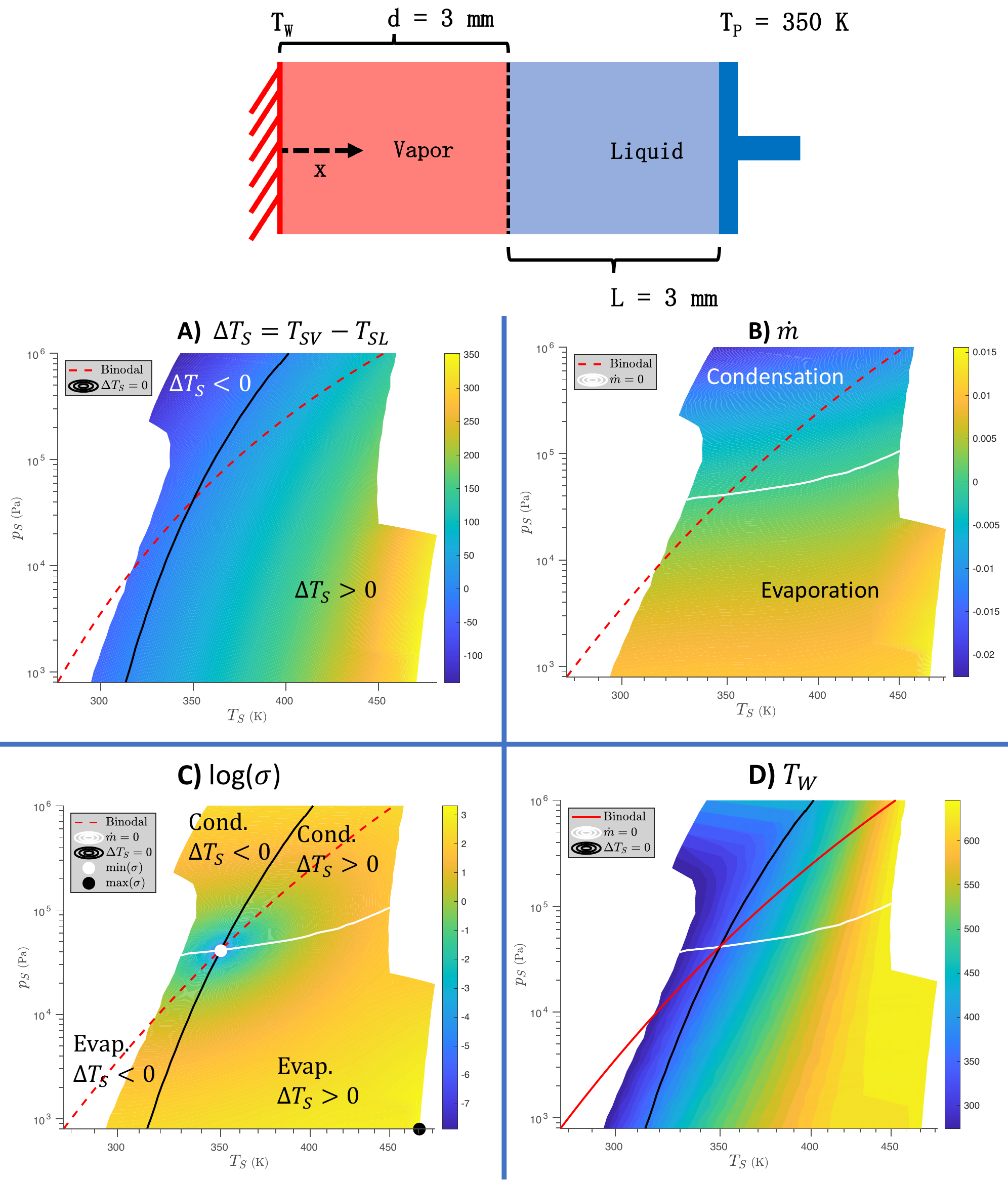}

\caption{The nonequilibrium properties at the two phase interface mapped onto the average interfacial pressure and temperature. The distance of the interface from the wall is 3 mm, corresponding to the width of the vapor domain. The distance of the interface to the piston is likewise 3 mm, corresponding to the width of the liquid domain. The piston cooling the liquid reservoir is held at a constant temperature of $350$ K, while the wall temperature is varied. \textbf{A}) The temperature jump at the liquid vapor interface is generally nonzero. The interface temperature is only continuous along the black contour. \textbf{B}) The mass flux due to phase change across the interface is positive for evaporation and negative for condensation. No phase change occurs along the white contour. \textbf{C}) The logarithm of the entropy production rate at each interface pressure and temperature shows that the global minimum is located at equilibrium on the binodal, where the continuous interface temperature and zero mass flux contours intersect. The point of intersection corresponds to a constant temperature profile equal to the far field piston temperature $350$ K in the liquid domain. \textbf{D}) The wall temperature $T_W$ is varied from $274$ K to $646$ K at different pressures to obtain the phase diagrams of the interfacial temperature jump, mass flux and entropy production rate. The vapor and liquid properties of water (thermal conductivities, latent heat of phase change, etc.) as a function of temperature were referenced from the IAPWS formulation \cite{2012Huber, 2002Wagner, 2012NIST}. }

\label{fig: 4contour}
\end{figure}

To gain a physical understanding of nonequilibrium phase change in a liquid-vapor system, we observe the interfacial temperature jump, mass flux and entropy production rate associated with the average pressure and temperature at the interface (Fig. \ref{fig: 4contour}). For this example, the widths of both the vapor and liquid domains are fixed at $3$ mm. The piston cooling the liquid reservoir is held at a constant temperature of $350$ K, while the wall temperature is varied for each value of pressure imposed on the system (Fig. \ref{fig: 4contour} \textbf{D}). Given these boundary conditions, the maximum entropy production rate is used to determine the interface temperatures $T^*_{SL}$ and $T^*_{SV}$, the average of which is plotted on the x axis. Similarly, the average pressure at the two phase interface is tabulated on the y axis. 

Fig. \ref{fig: 4contour} \textbf{A} overlays the interfacial temperature jump $\Delta T_S=T^*_{SV}-T^*_{SL}$ on the average pressure-temperature diagram at the interface. It is notable that interfacial temperature continuity is a special condition confined to a single contour, whereas the majority of the phase space is dominated by the existence of a temperature discontinuity. The sign of this jump, whether $T^*_{SV}>T^*_{SL}$ or vice versa, cannot in general be predicted by the equilibrium binodal curve. The assumption of temperature continuity is not generally reliable when the local interface rests in equilibrium.  

Another way to characterize the interface is to look at the mass flux due to phase change (Fig. \ref{fig: 4contour} \textbf{B}). The contour along which no phase change occurs is likewise ill predicted by the binodal in general; all three curves (binodal, $\dot{m}=0$, and $\Delta T_S=0$) only intersect at the point $T_W=T^*_{SV}=T^*_{SL}=T_P$, which reflects constant temperature profiles in both phases. This corresponds to the bulk system being in equilibrium (Fig. \ref{fig: 4contour} \textbf{C}). 

The two contours $\dot{m}=0$ and $\Delta T_S=0$ separate the interfacial phase space into four regions. In the top left and bottom right sectors, phase change conforms to our physical intuition around the binodal. That is, as temperature increases or pressure decreases past the coexistence curve, vapor becomes the bulk stable phase and vice versa. The top right and bottom left sectors characterize the metastable phases that are involved in processes like capillary condensation below and capillary evaporation above the binodal. Thus this entropic interface condition gives a complete description of possible phase change processes in nonequilibrium scenarios. 

Another issue of note is that Fig. \ref{fig: 4contour} \textbf{C} displays agreement with the minimum entropy production rate principle \cite{Martyushev_2006}. This principle says that over the relaxation time scale of a stationary nonequilibrium system with some thermodynamic forces fixed and others free, the thermodynamic fluxes in the system conjugate to those unfixed forces will disappear. This drives the system toward the minimum of the entropy production rate, which occurs at equilibrium. Indeed, the global minimum in the entropy production rate is associated with the equilibrium interfacial temperature and pressure, at the intersection of the two contours $\dot{m}=0$ and $\Delta T_S=0$. If the wall temperature is allowed to evolve over time from the initial condition (unfixed) rather than be held to a constant value, the system would eventually relax to this equilibrium state in which the temperature profiles in both phases are constant and equal.

On the shorter time scale, or if all thermodynamic forces are held constant, the associated thermodynamic fluxes adjust in order for the system to achieve the maximum entropy production rate for each specific, average interfacial pressure and temperature plotted in Fig. \ref{fig: 4contour} \textbf{C}. The minimum entropy production rate principle suggests that a stationary nonequilibrium system with sufficient degrees of freedom will tend toward minimum value of the entropy production rate over the relaxation time scale, whereas the maximum entropy production rate principle tells us that a nonequilibrium system on a shorter time period or under constant thermodynamic forcing will find the state corresponding to the maximum in the entropy production rate as the fluxes vary. Indeed, both principles can apply simultaneously, in that a stationary nonequilibrium system may approach the state associated with the minimum entropy production rate on a longer time scale by taking individual steps over a short time scale that maximize the entropy production rate at each step, while satisfying thermodynamic constraints and governing laws. 

Thus, the maximum entropy production rate allows us to accurately pinpoint the interfacial properties of a nonequilibrium system undergoing phase change in a thermodynamically consistent manner. It is the missing condition needed to close the two phase problem when phase change occurs across the interface. In addition, the minimum entropy production rate informs the trajectory of a stationary phase change system over longer time scales, within the permissible phase space set by the presence of fixed thermodynamic forces or fluxes.

\section{Conclusion}
The maximum entropy production rate at the interface closes the phase change problem by determining the interface temperature and velocity. The predictions from the proposed entropic interface condition well capture nanoscale temperatures and mass fluxes for liquid, vapor and solid phase change at the nanoscale. The condition also accurately predicts experimental data on interface temperature jumps and mass fluxes for different fluids under both turbulent and laminar flows at mesoscale. This agreement suggests that at most length and time scales, the interface properties are dominated by a deterministic thermodynamic principle (that of entropy production maximization) rather than stochastic or transient behavior which must be modeled probabilistically.

The maximum entropy principle can be used directly to design phase change systems to achieve desired mass fluxes or interface properties for the applications discussed prior. It can also be used to model nanoscale and mesoscale effects in continuum level simulations of multiphase flows, where the saturation temperature has been the standard approximation.

\section{Methods}
To gauge interface properties under nanoscale evaporation conditions as set up in Fig. \ref{fig: problem_setup}, molecular dynamics simulations were carried out with LAMMPS \cite{1995LAMMPS} software. A total of 32085 SPC/E molecules of liquid water were equilibrated at saturation temperature $T=373.15$ K in the canonical ensemble with constant pressure (1 atm) imposed by a piston constrained to move only in the direction orthogonal to the bottom surface \cite{2020Zhao}. The solid surface and piston were constructed using two graphene sheets with armchair lattice orientation, and the interaction between these planes and the SPC/E water molecules was governed by the 6-12 Lennard Jones pair potential with the depth of the potential well fixed at 0.05 kcal/mole \cite{2019Zhao}. After this equilibration step, the liquid water adjacent to the bottom surface was heated to a target temperature of $T=575$ K, whereas the liquid adjacent to the piston was held at constant, saturation temperature to simulate nonequilibrium heat transfer conditions \cite{2014heatconduction}. The lateral simulation box size in the plane parallel to the surface and piston was 8 nm by 8 nm. The perpendicular dimension varied as the vapor film thickness evolved in time.

\newpage

\section*{Author contributions}
N.A.P. and T.Y.Z. conceived and planned the research, performed the analyses, and wrote the manuscript.
\section*{Competing interests}
The authors have no competing financial interests or other interests that might be perceived to influence the results and/or discussion reported in this paper.
\section{Corresponding author}
To whom correspondence should be addressed. E-mail: n-patankar@northwestern.edu

\newpage

\bibliographystyle{unsrt}

\begin{thebibliography}{10}

\bibitem{2012MD}
Abdullah Alkhudhiri, Naif Darwish, and Nidal Hilal.
\newblock Membrane distillation: A comprehensive review.
\newblock {\em Desalination}, 287:2--18, 2012.

\bibitem{2018Sarbu}
Ioan Sarbu and Calin Sebarchievici.
\newblock A comprehensive review of thermal energy storage.
\newblock {\em Sustainability}, 10(1), 2018.

\bibitem{2019Simonelli}
Marco Simonelli, Nesma Aboulkhair, Mircea Rasa, Mark East, Christopher Tuck,
  Ricky Wildman, Otto Salomons, and Richard Hague.
\newblock Towards digital metal additive manufacturing via high-temperature
  drop-on-demand jetting.
\newblock {\em Additive Manufacturing}, 30:100930, 10 2019.

\bibitem{2010Burr}
Geoffrey~W. Burr, Matthew~J. Breitwisch, Michele Franceschini, Davide Garetto,
  Kailash Gopalakrishnan, Bryan Jackson, B{\"u}lent Kurdi, Chung Lam, Luis~A.
  Lastras, Alvaro Padilla, Bipin Rajendran, Simone Raoux, and Rohit~S. Shenoy.
\newblock Phase change memory technology.
\newblock {\em Journal of Vacuum Science \& Technology B}, 28(2):223--262,
  2019/12/31 2010.

\bibitem{1998Juric}
Damir Juric and Gr{\'e}tar Tryggvason.
\newblock Computations of boiling flows.
\newblock {\em International Journal of Multiphase Flow}, 24(3):387--410, 1998.

\bibitem{2005Dhir}
Vijay~K. Dhir.
\newblock {Mechanistic Prediction of Nucleate Boiling Heat Transfer--Achievable
  or a Hopeless Task?}
\newblock {\em Journal of Heat Transfer}, 128(1):1--12, 10 2005.

\bibitem{2007Badam}
V.~K. Badam, V.~Kumar, F.~Durst, and K.~Danov.
\newblock Experimental and theoretical investigations on interfacial
  temperature jumps during evaporation.
\newblock {\em Experimental Thermal and Fluid Science}, 32(1):276--292, 2007.

\bibitem{2017Gatapova}
Elizaveta~Ya. Gatapova, Irina~A. Graur, Oleg~A. Kabov, Vladimir~M. Aniskin,
  Maxim~A. Filipenko, Felix Sharipov, and Loun{\`e}s Tadrist.
\newblock The temperature jump at water --air interface during evaporation.
\newblock {\em International Journal of Heat and Mass Transfer}, 104:800--812,
  2017.

\bibitem{1991Tanasawa}
Ichiro Tanasawa, James~P. Hartnett, Thomas~F. Irvine, and Young~I. Cho.
\newblock {\em Advances in Condensation Heat Transfer}, volume~21, pages
  55--139.
\newblock Elsevier, 1991.

\bibitem{2001Ward}
C.~A. Ward and D.~Stanga.
\newblock Interfacial conditions during evaporation or condensation of water.
\newblock {\em Phys. Rev. E}, 64:051509, Oct 2001.

\bibitem{2003Dewar}
Roderick Dewar.
\newblock Information theory explanation of the fluctuation theorem, maximum
  entropy production and self-organized criticality in non-equilibrium
  stationary states.
\newblock 36(3):631--641, 2003.

\bibitem{2010Niven}
Robert~K Niven.
\newblock Minimization of a free-energy-like potential for non-equilibrium flow
  systems at steady state.
\newblock {\em Philosophical transactions of the Royal Society of London.
  Series B, Biological sciences}, 365(1545):1323--1331, 05 2010.

\bibitem{1992Carey}
Berkeley CA (United States). Dept. of Mechanical~Engineering) Carey, V. P.
  (California~Univ.
\newblock {\em Liquid-vapor phase-change phenomena}.
\newblock New York, NY (United States); Hemisphere Publishing, United States,
  1992.

\bibitem{1974Delhaye}
J.~M. Delhaye.
\newblock Jump conditions and entropy sources in two-phase systems. local
  instant formulation.
\newblock {\em International Journal of Multiphase Flow}, 1(3):395--409, 1974.

\bibitem{1973Meinhold}
L.~Meinhold-Heerlein.
\newblock Surface conditions for the liquid-vapor system, taking into account
  entropy production caused by mass and energy transport across the interface.
\newblock {\em Phys. Rev. A}, 8:2574--2585, Nov 1973.

\bibitem{2020Zhao}
Tom~Y. Zhao and Neelesh~A. Patankar.
\newblock The thermo-wetting instability driving leidenfrost film collapse.
\newblock {\em Proceedings of the National Academy of Sciences}, page
  201917868, 05 2020.

\bibitem{2012NIST}
E.W. Lemmon, M.O. McLinden, and D.G. Friend.
\newblock {\em NIST Chemistry WebBook, NIST Standard Reference Database Number
  69, Eds. P.J. Linstrom and W.G. Mallard}.
\newblock Gaithersburg MD, 2012.

\bibitem{2006MEP}
L.~M. Martyushev and V.~D. Seleznev.
\newblock Maximum entropy production principle in physics, chemistry and
  biology.
\newblock {\em Physics Reports}, 426(1):1--45, 2006.

\bibitem{2010Martyushev}
Leonid~M. Martyushev.
\newblock The maximum entropy production principle: two basic questions.
\newblock {\em Philosophical Transactions of the Royal Society B: Biological
  Sciences}, 365(1545):1333--1334, 2020/07/24 2010.

\bibitem{1978Paltridge}
G.~W. Paltridge.
\newblock The steady-state format of global climate.
\newblock {\em Quarterly Journal of the Royal Meteorological Society},
  104(442):927--945, 2020/07/24 1978.

\bibitem{2010Vladislav}
Vladislav~V. Levchenko, Ronan Fleming, Hong Qian, and Daniel~A. Beard.
\newblock An annotated english translation of `kinetics of stationary
  reactions' [m. i. temkin, dolk. akad. nauk sssr. 152, 156 (1963)], 2010.

\bibitem{2017Endres}
Robert~G. Endres.
\newblock Entropy production selects nonequilibrium states in multistable
  systems.
\newblock {\em Scientific Reports}, 7(1):14437, 2017.

\bibitem{Holten_2013}
Vincent Holten, David~T. Limmer, Valeria Molinero, and Mikhail~A. Anisimov.
\newblock Nature of the anomalies in the supercooled liquid state of the mw
  model of water.
\newblock {\em The Journal of Chemical Physics}, 138(17):174501, May 2013.

\bibitem{2017Wang}
Tianbao Wang and Min Chen.
\newblock Determining interface temperature during rapid freezing of
  supercooled water.
\newblock {\em Journal of Aircraft}, 55(3):1269--1275, 2020/06/04 2017.

\bibitem{1999Ward2}
G.~Fang and C.~A. Ward.
\newblock Temperature measured close to the interface of an evaporating liquid.
\newblock {\em Phys. Rev. E}, 59:417--428, Jan 1999.

\bibitem{2008Ward}
Fei Duan, C.~A. Ward, V.~K. Badam, and F.~Durst.
\newblock Role of molecular phonons and interfacial-temperature discontinuities
  in water evaporation.
\newblock {\em Phys. Rev. E}, 78:041130, Oct 2008.

\bibitem{1999Ward}
G.~Fang and C.~A. Ward.
\newblock Examination of the statistical rate theory expression for liquid
  evaporation rates.
\newblock {\em Phys. Rev. E}, 59:441--453, Jan 1999.

\bibitem{2004Ward}
C~A Ward and Fei Duan.
\newblock Turbulent transition of thermocapillary flow induced by water
  evaporation.
\newblock {\em Phys Rev E Stat Nonlin Soft Matter Phys}, 69(5 Pt 2):056308, May
  2004.

\bibitem{2012Huber}
M.~L. Huber, R.~A. Perkins, D.~G. Friend, J.~V. Sengers, M.~J. Assael, I.~N.
  Metaxa, K.~Miyagawa, R.~Hellmann, and E.~Vogel.
\newblock New international formulation for the thermal conductivity of h2o.
\newblock {\em Journal of Physical and Chemical Reference Data}, 41(3):033102,
  2020/06/28 2012.

\bibitem{2002Wagner}
W.~Wagner and A.~Pru{\ss}.
\newblock The iapws formulation 1995 for the thermodynamic properties of
  ordinary water substance for general and scientific use.
\newblock {\em Journal of Physical and Chemical Reference Data},
  31(2):387--535, 2020/06/28 2002.

\bibitem{Martyushev_2006}
L~M Martyushev, A~S Nazarova, and V~D Seleznev.
\newblock On the problem of the minimum entropy production in the
  nonequilibrium stationary state.
\newblock {\em Journal of Physics A: Mathematical and Theoretical},
  40(3):371--380, dec 2006.

\bibitem{1995LAMMPS}
Steve Plimpton.
\newblock Fast parallel algorithms for short-range molecular dynamics.
\newblock {\em Journal of Computational Physics}, 117(1):1 -- 19, 1995.

\bibitem{2019Zhao}
Tom~Y. Zhao, Paul~R. Jones, and Neelesh~A. Patankar.
\newblock Thermodynamics of sustaining liquid water within rough icephobic
  surfaces to achieve ultra-low ice adhesion.
\newblock {\em Scientific Reports}, 9(1):258, 2019.

\bibitem{2014heatconduction}
Junichiro Shiomi.
\newblock Nonequilibrium molecular dynamics methods for lattice heat conduction
  calculations.
\newblock {\em Annual Review of Heat Transfer}, 17:177--203, 01 2014.

\end{thebibliography}

\end{document}